\documentclass[a4paper,11pt]{article}
\pdfoutput=1 
   
\usepackage{jheppub} 
\usepackage{url}
\usepackage{caption}
\usepackage{subcaption}

\usepackage[latin9]{inputenc}
\usepackage{float}
\usepackage{amsmath}
\usepackage{amssymb}
\usepackage{graphicx}
\usepackage{esint}
\usepackage{hyperref}
\usepackage{comment}
\usepackage{color}
\usepackage{microtype}
\usepackage{cleveref}
\usepackage{breakurl}
\usepackage{bbm}
\newcommand{\be}{\begin{equation}}
\newcommand{\ee}{\end{equation}}
\newcommand{\ben}{\begin{displaymath}}
\newcommand{\een}{\end{displaymath}}
\newcommand{\bea}{\begin{eqnarray}}
\newcommand{\eea}{\end{eqnarray}}
\def\K{K{\"a}hler}
   \newcommand{\rf}[1]{(\ref{#1})}

\def\E{{$E_{7(7)}$}}

\def\be{\begin{equation}}
\def\ee{\end{equation}}
\def\bea{\begin{eqnarray}}
\def\eea{\end{eqnarray}}
\def\bit{\begin{itemize}}
\def\eit{\end{itemize}}

\def\Tr{{\rm Tr}}

\newcommand{\cN}{\mathcal{N}}

\def\E{{$E_{7(7)}$}}

 \makeatletter

\allowdisplaybreaks

\makeatother

\makeatletter
\DeclareRobustCommand{\rcite}[1]{%
  \rcite@aux#1,\@nil{#1}%
}
\def\rcite@aux#1,#2\@nil#3{%
  \if\relax#2\relax
    Ref.~\cite{#3}%
  \else
    Refs.~\cite{#3}%
  \fi
}
\makeatother

\hypersetup{
    colorlinks = true,
    citecolor = {blue},
    linkcolor = {blue},
    urlcolor = {blue},
}

 \title{\rm { \LARGE \bf        M-theory, Black Holes and  Cosmology }}

\author[a]{Renata Kallosh,}

\affiliation[a]{Stanford Institute for Theoretical Physics and Department of Physics,\\ Stanford University, Stanford, CA 94305, USA}

\emailAdd{kallosh@stanford.edu}

\notoc

\abstract{  This paper is dedicated to M. Duff on the occasion of his 70th birthday. I discuss some issues of M-theory/string theory/supergravity  closely related to Mike's interests.  I describe a relation between STU black hole entropy, Cayley hyperdeterminant,  Bhargava cube and  a 3-qubit  Alice, Bob, Charlie triality symmetry. I shortly  describe my recent work with  Gunaydin,  Linde, Yamada on M-theory cosmology \cite{Gunaydin:2020ric},  inspired by the work of  Duff with  Ferrara and  Borsten,  Levay, Marrani et al.  Here we have  7-qubits, a  
party  including Alice, Bob, Charlie, Daisy, Emma, Fred, George. Octonions and   Hamming  error correcting codes are at the base of these models. They  lead to  7 benchmark targets of future CMB missions looking for primordial gravitational wave from inflation.
I also show  puzzling relations between the fermion mass eigenvalues in these cosmological models, exceptional Jordan eigenvalue problem,  and  black hole entropy. The  symmetry of our cosmological models is illustrated by beautiful pictures of a Coxeter projection of the root system of E7.}

\begin{document}

\maketitle

   \newpage

\tableofcontents{}

  \parskip 8pt

 \section{Introduction}\label{intro}

Inspiring ideas of Mike Duff  have influenced my work over decades. Here I would like to present two aspects of it,
both rooted in Mike's and in my work from long time ago, where there  is also a very recent progress.

 The first story is about the  black holes attractors \cite{Ferrara:1996dd} in $\cN=2$ 4d supergravity, originating from M-theory/string theory/11d supergravity. These black holes have interesting properties which were initially understood in \cite{Duff:1995sm} and  \cite{Behrndt:1996hu} where the STU black holes with string theory triality symmetry were described. These were followed by \cite{Duff:2006uz} and \cite{Kallosh:2006zs} where the STU black holes and their entropies were related to quantum information theory. In these papers the relation between quantum entanglement in a 3-qubit system, Alice, Bob and Charlie, and and 3-moduli STU black holes was discussed. 
 
 The recent stage of this story has to do with a renewed interest in mathematical aspects of black holes in string theory/supergravity as studied in \cite{Benjamin:2018mdo, Gunaydin:2019xxl,Banerjee:2020lxj,Borsten:2020nqi}. The  relation between STU black holes and Bhargava cube was observed and discussed earlier  in \cite{Borsten:2008zz,Borsten:2008wd}. 
  We will add to the recent advances in all these papers the analysis of the triality  symmetry, which exists for these black holes in addition to the well known and well studied U-duality $[SL(2,\mathbb{Z})]^3$ symmetry. Basically triality symmetry  is a statement that Alice, Bob and Charlie are on equal footing. The aspects of Bhargava cube related to properties of the Cayley hyperdeterminant   will be discussed  here. We will clarify the concept of equivalence of  black holes with the same entropy with 
 U-duality symmetry $[SL(2, \mathbb{Z})]^3 \ltimes S_3$.
 
It was  noticed in \cite{Kallosh:2006zs} that the black holes in $\cN=8$ 4d supergravity can be brought to a canonical basis. Their entropy formula  defined in general by  56 charges in the quartic Cartan-Cremmer-Julia \E\,  invariant, in the canonical basis
 depends only on 8 charges and coincides  with the Cayley hyperdeterminant  defining the  STU black holes area of the horizon/entropy.  In  the Bhargava cube terminology this $[SL(2,\mathbb{Z})]^3$ invariant is  a discriminant of the associated  binary quadratic forms.
 
The $[SL(2, \mathbb{Z})]^3 \ltimes S_3$  symmetry of the  Cayley hyperdeterminant/Bhargava cube is 
 also a symmetry following from the \K\, potential which is given by
\be
K_{3{\rm mod}}= - \sum_{i=1}^3 \log\left(  T^i + \overline{T}^i\right)\, .
\label{STUK}\ee 
 STU black holes can be associated with M-theory first truncated to 7 moduli, $T^i$,  $i=1,\dots , 7$ 
with  $[SL(2,\mathbb{Z})]^7$ and $S_7$ symmetry and the \K\, potential  given by
\be
K_{7{\rm mod}}= - \sum_{i=1}^7 \log\left(  T^i + \overline{T}^i\right)\, .
\label{7_K}\ee 
 When 4 of the 7 moduli   are  truncated we have the remaining 
 $[SL(2,\mathbb{Z})]^3$ duality as well as triality permutation symmetry $S_3$, and we recover the kinetic term of  $\cN=2$ supergravity STU model. A detailed derivation of the STU model from string theory/10d supergravity was performed in \cite{Duff:1995sm}.
 
The second story of this paper is about  the new ideas in cosmology based on 7-moduli model of M-theory compactified on a manifold with $G_2$ holonomy and with  $[SL(2,\mathbb{Z})]^7$ symmetry and \K\, potential in eq. \rf{7_K}. M. Duff was the first to point out  in \cite{Awada:1982pk}  that the maximal supersymmetry of M-theory is spontaneously broken down to $\cN=1$ supersymmetry in 4d when compactified on a manifold with $G_2$ holonomy. More recently  11d M-theory/supergravity  compactified on  a twisted 7-tori   with holonomy  group  $G_2$ was investigated in
 \cite{DallAgata:2005zlf}.

\begin{figure}[H]
\centering
\includegraphics[scale=1]{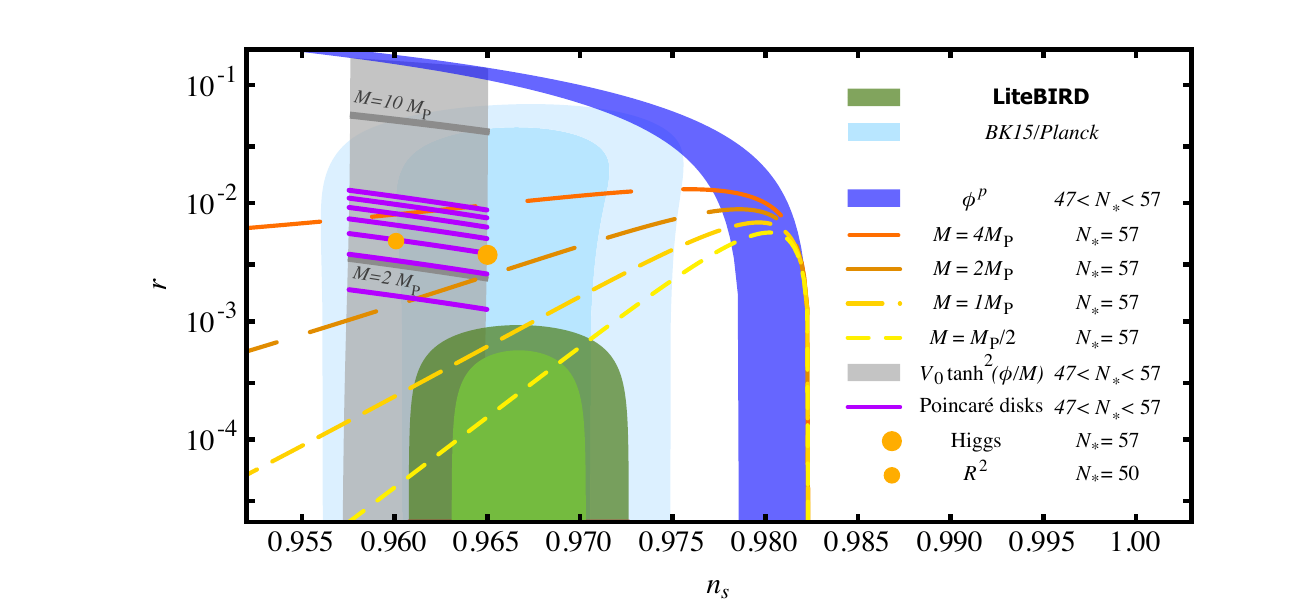}
\caption{\footnotesize  This is a figure A.2 from the Astro2020 APC White Paper~\href{https://ui.adsabs.harvard.edu/abs/2019BAAS...51g.286L/abstract}{LiteBIRD: an all-sky cosmic microwave background probe of inflation} with a forecast of Litebird constraints in the $n_s$ - $r$ plane~\cite{2019BAAS...51g.286L}.
The 7  purple lines in the figure, were derived most recently in \cite{Gunaydin:2020ric}  using M-theory compactified on $G_2$, octonions, Fano planes and error correcting codes.
}
\label{fig:LB}
\end{figure}

 During the last few years I studied the issues in cosmology initiated by discussions with S. Ferrara  which resulted in our paper \cite{Ferrara:2016fwe}.  This work, in turn, originated from S. Ferrara's work with M. Duff and his collaborators \cite{Duff:2006ue,Levay:2006pt,Borsten:2008wd,Levay:2010hna,Borsten:2012fx}. One of the central ideas in all these studies is based on the fact that \E\,$(\mathbb{R})$  symmetry of $\cN=8$ 4d supergravity has a subgroup $[SL(2,\mathbb{R})]^7$. For the discrete subgroups this becomes a following relation
 \be
 E_{7(7)} (\mathbb{Z}))\supset [SL(2,\mathbb{Z})\Big]^7\, . 
 \ee
 When the relevant cosmological models were constructed in \cite{Ferrara:2016fwe,Kallosh:2017ced,Kallosh:2017wnt,Kallosh:2019hzo},  7 targets for early universe future searches of gravitational waves from inflation were proposed. These are shown here in Fig. \ref{fig:LB} by 7 purple lines.
 
 The theoretical  underpinning of the cosmological models in \cite{Ferrara:2016fwe,Kallosh:2017ced,Kallosh:2017wnt,Kallosh:2019hzo} was very recently proposed in 
  my paper  \cite{Gunaydin:2020ric} with M. Gunaydin, A. Linde, Y. Yamada. The entangled 7-qubit system corresponds to 7  parties: Alice, Bob, Charlie, Daisy, Emma, Fred and George,  and it is related to 7 imaginary units of octonions.
   
M. Duff had a long and deep appreciation of the fact that  there are  four normed division algebras: the real numbers ($\mathbb{R}$), complex numbers ($\mathbb{C}$), quaternions ($\mathbb{H}$), and octonions ($\mathbb{O}$).  He and his collaborators have developed many new aspects of the relations between  octonions and physics, see for example  \cite{Borsten:2008wd}. I will show here how octonions,  Fano planes and error correcting Hamming (7,4) codes help to build cosmological models which will be tested by future cosmological observations. 

I will also show that the mass eigenvalues of heavy scalars  in cosmological models in  \cite{Gunaydin:2020ric} described by a pair of cubic equations
$x^3-7x-7=0, \, y^3-7x-7=0$ have a particular relation to  exceptional Jordan\footnote{Studies of octonions and Jordan algebras are based on the work of H. Freudenthal  in `Oktaven, Ausnahmegruppen und Oktavengeometrie', Mathematisch Instituut der Rijksuniversiteit te Utrecht, 1951.}
  eigenvalue problem \cite{Gunaydin:1978jq,Dundarer:1983fe,Ferrara:1997uz,Ferrara:2006xx,Dray:1999xe}. There is an interesting connection between the product of the mass eigenvalues of fermions in cosmological models and the entropy of the  STU black holes. Both correspond to a determinant of a certain relevant in each case Jordan matrix.

Another interesting feature of our cosmological models \cite{Gunaydin:2020ric}  is the symmetry of the  fermion mass matrix 
 at Minkowski vacua. It is  invariant under the $O( 7)$  symmetry and its subgroups. The discrete subgroup of it is the Weyl group $W(E_7)$. We show the Coxeter plane of the   root system of $W(E_7)$ in Figs. \ref{fig:E7roots1}, \ref{fig:E7roots2}. When one imposes the invariance of the octonion algebra on the transformations one obtains a finite subgroup of $G_2$,  the adjoint Chevalley group $G_2(2)$ of order 12,096 as discussed in
\cite{Koca:1988xe,Karsch:1989gj,Koca:2005fn,Anastasiou:2015ena}.  This is interesting since it is expected that neutrino physics will require an extension of the standard model. Some of these extensions might include  discrete subgroups of $G_2$, see  for example \cite{Ramond:2020dgm,Perez:2020nqq}.

Thus, both of these stories, STU black holes and M-theory cosmology 7-moduli models have interesting connection to $E_7$ symmetry.  I would like to notice here that  the current status of 4d $\cN=8$ supergravity and its perturbative UV behavior remain puzzling. Some heroic efforts were made by Z. Bern et al  in amplitude loop computations, see the review \cite{Bern:2019prr}. They  have shown that maximal supergravity behaves in UV much better than expected. It was suggested in \cite{Beisert:2010jx,Kallosh:2018wzz,Gunaydin:2018kdz}, 
that  \E\,  symmetry together with maximal supersymmetry of perturbative  maximal supergravity in 4d might explain the cancellation of UV infinities observed in `theoretical experiments' as described in   \cite{Bern:2019prr}. It would be very interesting to learn more about these exceptional symmetries and their role in physics.

 \section{STU black holes, triality and the Bhargava cube}
 
\begin{figure}[h!]
\centerline{ \epsfxsize 3 in\epsfbox{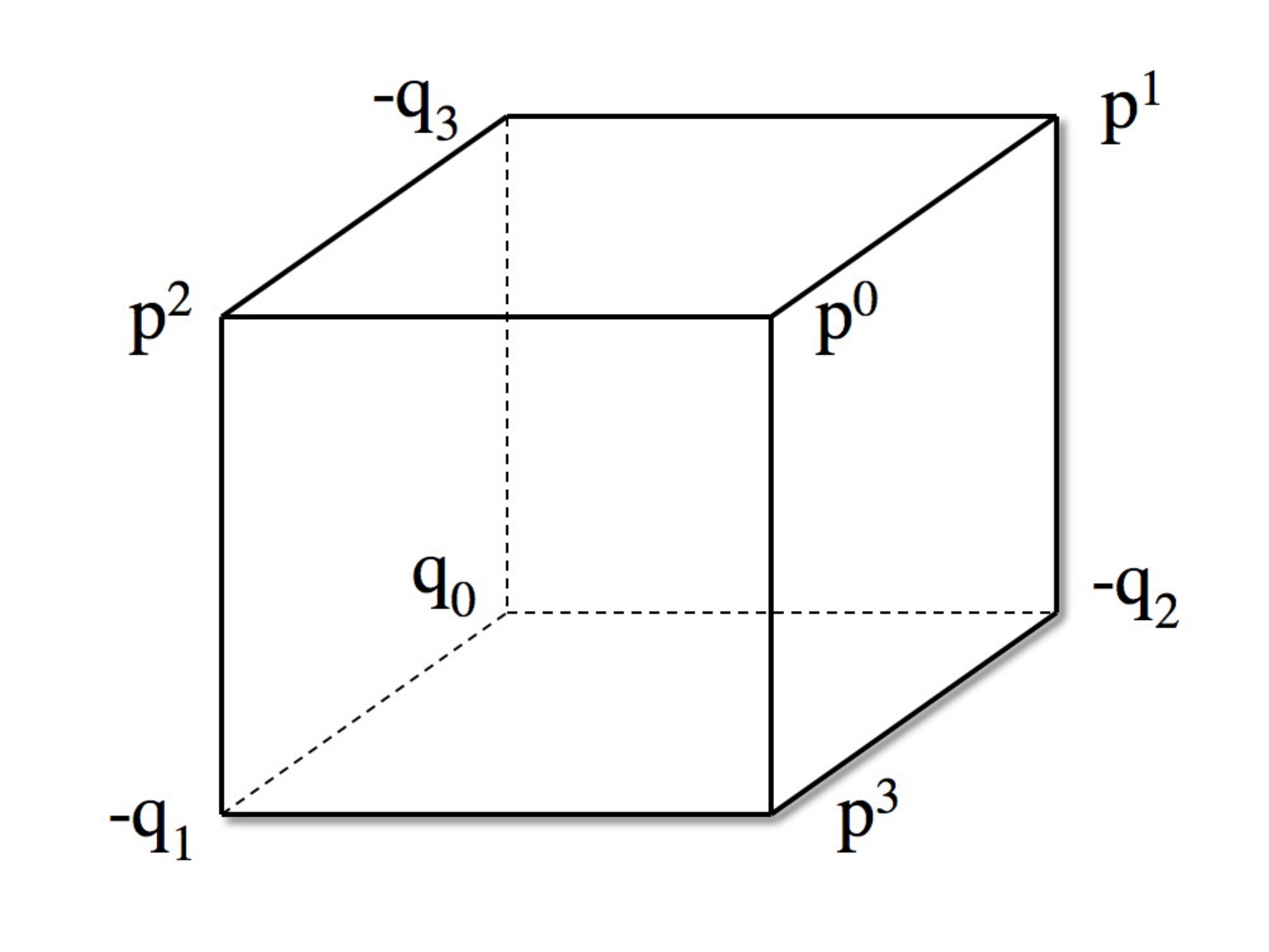}}
\caption{The 2x2x2 hypermatrix  corresponding to supergravity black holes  given in Fig. 2 of \cite{Kallosh:2006zs}. It represents the   STU black hole solution in \cite{Behrndt:1996hu} with 8 charges $p^\Lambda=p^0, p^1, p^2, p^3$ and $q_\Lambda=q_0, q_1, q_2, q_3$ where 
3 moduli  $z^1, z^2, z^3$ at the black hole horizon are functions of these charges. }
\label{stu2cube}
\end{figure}

 A significant effort was dedicated over the years to understand the properties of black holes in M-theory/string theory 
 /supergravity. The STU black holes are sufficiently simple, there are exact analytic solutions in classical $\cN=2$ supergravity with the prepotential 
\be
F= {d_{ijk} X^i X^j X^k \over X^0}= { X^1 X^2 X^3 \over X^0}
\label{prep} \ee in the so-called double extreme approximation, when the values of 3 moduli  $z^i= {X^i\over X^0}$ near the horizon are the same as the ones far away from the black hole, $z^i|_{\inf}=z^i|_{hor}$. The solution depends on 8 charges, 4 electric and 4 magnetic. The area of the horizon/the entropy of these  black holes was computed in \cite{Behrndt:1996hu} in terms of the 8 black hole charges $(p^\Lambda, q_{\Lambda})$, $\Lambda=0,1,2,3$, shown as corners of the 2x2x2 hypermatrix in Fig. \ref{stu2cube}. 
\noindent The entropy is the function of charges
 \be
{S\over \pi}= \left( W(p^\Lambda,q_\Lambda)\right)^{1/2} \ ,
\ee
where
\bea \label{ww}
 W(p^\Lambda ,q_\Lambda) =-{(p\cdot q)}^2&+&4\bigl (
(p^1q_1)(p^2q_2)+(p^1q_1)(p^3q_3)+(p^3q_3)(p^2q_2)\bigr )\cr
\cr
&-& 4 p^0 q_1 q_2 q_3 + 4q_0 p^1 p^2 p^3 
\eea
and
\begin{eqnarray}
p\cdot q \equiv( p^0 q_0) + ( p^1 q_1)  +( p^2 q_2)+ ( p^3 q_3)\ .
\end{eqnarray}\label{aaa}
The function $ W(p^\Lambda ,q_\Lambda)$ is manifestly symmetric under
transformations: 
\be
p^1\leftrightarrow p^2 \leftrightarrow p^3\, ,  \qquad 
  q_1\leftrightarrow q_2 \leftrightarrow q_3 \, . 
 \label{perm} \ee 
Under $[SL(2,\mathbb{Z})]^3$ transformations the charges and the moduli transform but the entropy is invariant.

\begin{figure}[h!]
\centerline{ \epsfxsize 2.3 in\epsfbox{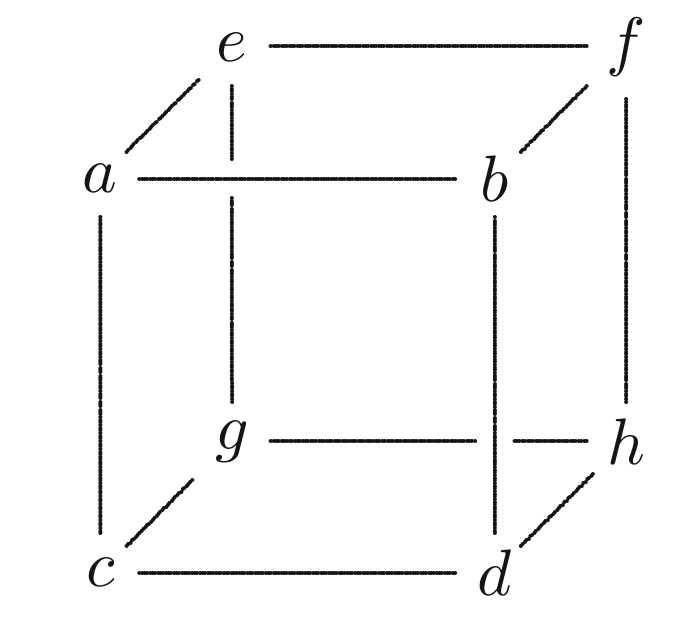}}
\caption{The Bhargava cube. }
\label{Bhargava}
\end{figure}

The values of the 3 complex moduli near the horizon, for each $i=1,2,3$, were computed in \cite{Behrndt:1996hu}
\be
z^i= {B^i\over 2 A_i} \mp i {\sqrt {B^i- 4A_i C^i} \over 2 A_i}
\label{mod}\ee
where for each $i=1,2,3$
\be
A_i= p^0 q_i- 3 d_{ijk} p^j p^k \qquad B^i= p\cdot q- 2 p^iq_i \qquad C^i= -(p^i q_0 + 3d^{ijk} q_jq_k)
\ee  
and 
\be
W= -D=  B^1- 4A_1 C^1 = B^2- 4A_2 C^2=B^3- 4A_3 C^3\, . 
\ee 
 It was  pointed out in \cite{Duff:2006uz} that the classical expression for the entropy of the STU black holes $ W(p^\Lambda ,q_\Lambda)$ (\ref{ww})  can be represented in a very beautiful form:
\be
S^{\rm BPS}=  \pi  \sqrt W  = {\pi\over 2} \sqrt{-\rm Det ~\psi}\ , \qquad  \rm Det ~\psi  <0 \ , 
\ee
where $\rm Det ~\psi$ is the Cayley's hyperdeterminant of the  vector with components $\psi_{ijk}$, constructed in 1845. The dictionary between 8 charges  $p^\Lambda$ and $q_\Lambda$ and components of $\psi_{ijk}$ is the  following:
\be
\begin{array}{cccccccc}p^0 & p^1 & p^2 & p^3 & q_0 & q_1 & q_2 & q_3 
\cr
\\\psi_{000} & - \psi_{001} & - \psi_{010} & -\psi_{100} &  \psi_{111} & \psi_{110} & \psi_{101} & \psi_{011}\end{array}
\label{dic}\ee
Cayley hyperdeterminant of the 2x2x2 hypermatrix $\psi_{ijk}$ is defined as follows
 \bea
{\rm Det}~\psi&=-&\frac{1}{2}\epsilon^{ii^\prime}\epsilon^{jj^\prime}\epsilon^{kk^\prime}
\epsilon^{mm^\prime}\epsilon^{nn^\prime}\epsilon^{pp^\prime}\psi_{ijk}\psi_{i^\prime j^\prime m}
\psi_{npk^\prime}\psi_{n^\prime p^\prime m^\prime} \, . 
        \label{cayley}
\eea
The new aspect of the STU black holes associated with Bhargava cube developed in  \cite{Benjamin:2018mdo, Gunaydin:2019xxl,Banerjee:2020lxj,Borsten:2020nqi} is the following. 
It is possible to attach a triple of quadratic forms 
\be
A_i x^2 + B_i xy+C_i y^2
\ee
of the same discriminant $D = B_i^2 - 4A_iC_i$
 to a cube, with the corners given by an octuple $a, b, c, d, e, f, g, h$. We show this cube in Fig. \ref{Bhargava}. Even when only 2 of the forms are available, one can construct the third one as well as the Bhargava cube.
The dictionary between 8 black hole charges in Fig. \ref{stu2cube} and Bhargava octuple in Fig. \ref{Bhargava} is 
\be
\begin{array}{cccccccc}p^2 & p^0 & -q_1 & p^3 & -q_3 & p^1 & q_0 & -q_2 \\
\cr
a & b & c & d & e & f & g & h\end{array}
\label{dic1}\ee
The cube has a 3-way slicing: up-down, left-right, front-back,  and many interesting properties.
 The discriminant of the cube is given by the following expression.
\bea
D_{Bha} = &&a^2h^2+ b^2g^2+c^2f^2+d^2e^2\cr
\cr 
&&-2 (abgh+cdef+acfh+bdeg+aedh+bfcg)\cr
\cr
&&+4 (adfg+bceh)\, . 
\label{Bha}\eea
Using eqs. \rf{ww}, \rf{Bha} and the dictionary \rf{dic} we see that
\be
D_{Bha} = -W\, . 
\ee
 The action of modular groups $[SL(2,\mathbb{Z})]^3$ on the Bhargava cube was studied in detail in mathematical literature and recently applied in the context of STU black holes in \cite{Benjamin:2018mdo, Gunaydin:2019xxl,Banerjee:2020lxj,Borsten:2020nqi}. However, the permutations symmetry of the discriminant of the
 Bhargava cube was not yet revealed in most of these studies\footnote{Examples with triality symmetry were given  in  \cite{Borsten:2020nqi}, here we  discuss a general case of triality symmetry in the context of Bhargava cube. }. 
 Namely, the 3 permutation permutation symmetries  in black hole solutions  which preserve the entropy and reflect the symmetry between 3 moduli $z^i$ at the horizon and the relevant charges, are
  \bea
z^1 \leftrightarrow z^2 \quad : \quad &&p^1\leftrightarrow p^2 \qquad q_1\leftrightarrow q_2 \qquad \cr
\cr
z^2 \leftrightarrow z^3 \quad : \quad &&p^2\leftrightarrow p^3 \qquad q_2\leftrightarrow q_3\cr
\cr
z^3 \leftrightarrow z^1 \quad : \quad &&p^3\leftrightarrow p^1 \qquad q_3\leftrightarrow q_1
\label{permSymBH} \eea 
 
 Therefore the 3 symmetries of the  discriminant of the Bhargava cube  which reflect the corresponding black hole symmetries are

 \bea
&&f\leftrightarrow a \qquad c\leftrightarrow h\cr
\cr
&&a\leftrightarrow d \qquad h\leftrightarrow e\cr
\cr
&&d\leftrightarrow f \qquad e\leftrightarrow c
\label{permSym} \eea
We show them by red diagonal lines in Fig. \ref{Bhargava3Perm}:
 \begin{figure}[h!]
\centerline{ \epsfxsize 6 in\epsfbox{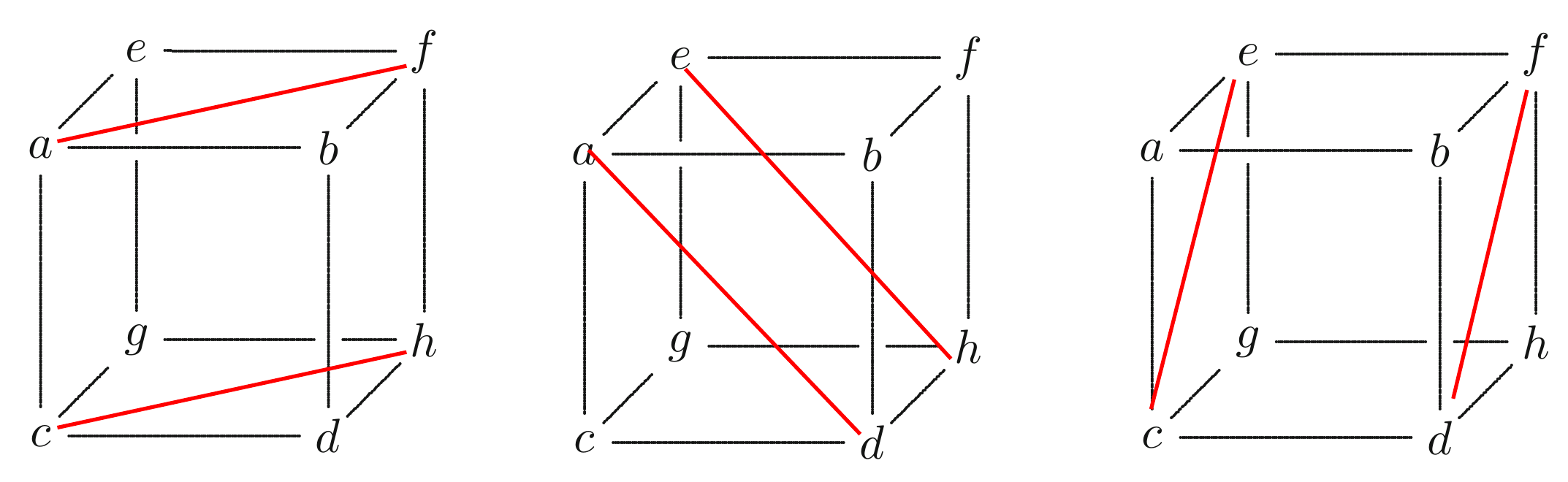}}
\caption{Three permutation symmetries of the discriminant of the  Bhargava cube $D_{Bha}$, according to   eq. \rf{permSym}}
\label{Bhargava3Perm}
\end{figure} 
\subsection{The issue of black hole equivalence}
Supersymmetric STU black holes are defined by their entropy as well as by the values of the 3 moduli near the horizon. In the basis where all 3 moduli $z^i$ are on equal footing in the prepotential given in eq. \rf{prep}, entropy is shown in eq.  \rf{ww} and the values of the moduli $z^i$ are given in eq. \rf{mod}. The $\cN=2$ supergravity in this basis  and the black hole solution both have this symmetry. The symmetry of solutions is presented in eq. \rf{permSymBH}.
 
The permutation symmetry for black holes, a triality symmetry, is important when the physical question is asked: what kind of STU black holes are equivalent? It is known that the entropy might be the same for different set of 8 changes, for example  $(p^\Lambda, q_\Lambda)$ and $\Big ((p^\Lambda)' , (q_\Lambda)'\Big)$. However, some of these 8 charges  can be related to each other by a  U-duality symmetry $[SL(2, \mathbb{Z})]^3 \ltimes S_3$ transformation. In such case, these  two sets of 8 charges belong to the same U-duality orbit. If however, they are not  related to each other by an $[SL(2, \mathbb{Z})]^3 \ltimes S_3$  transformation, they  belong to different orbits. 

There is a significant progress in understanding the discrete properties of  Bhargava cube which may be useful in the context of string theory counting of states associated with supersymmetric black holes with integer charges. To use these properties it would be nice to take into account systematically also triality symmetry $S_3$ of the discriminant of the Bhargava cube, in addition to modular $[SL(2, \mathbb{Z})]^3$ symmetry which was already studied extensively.

In the basis $a,b,c,d,e,f,g,h$ which is standard in Bhargava cube literature, this $S_3$ symmetry \rf {permSym} of the discriminant in eq. \rf{Bha} is not obvious since it does not appear to be related to a supergravity $S_3$ invariant prepotential \rf{prep}.  However, it is present there. The metric, and therefore the entropy of the STU black hole solution is U-duality invariant.  The $S_3$ symmetry is therefore manifest  in eq. \rf{ww} since it follows from the triality invariant prepotential.

 \section{M-theory cosmology, octonions and error correcting codes}
 A short summary of the recent paper \cite{Gunaydin:2020ric} suitable for this set up is the following. We have proposed an expression for the effective $\cN=1$ 4d supergravity following from M-theory/11d supergravity compactified on a manifold with $G_2$ structure.  
Starting with general type $G_2$-structure manifolds one finds Minkowski vacua only in cases the twisted 7-tori are $G_2$-holonomy manifolds. Here again it was a crucial  early insight  of M. Duff that 
  the maximal supersymmetry of M-theory is spontaneously broken by compactification to minimal 
   $\cN=1$ supersymmetry in 4d \cite{Awada:1982pk} when the compactification manifold has a $G_2$ holonomy.

Our choice of the superpotential  is based on a split of the 7-qubit system,  Alice, Bob, Charlie, Daisy, Emma, Fred, George, into 3-qubits and 4-qubits. The 3-qubits codify the multiplication table of octonions, there are 7 associated triads there. The 7 complimentary  4-qubits define our superpotential. The automorphism group of octonions is $G_2$, therefore it is natural to define the superpotentials using octonions.

The effective $\cN=1$ 4d supergravity following from M-theory/11d supergravity is defined as follows. The \K\, potential is given in eq. \rf{7_K}. The superpotential in general is given by a sum over 7 the 4-qubits $\{ijkl\}$ of the form
\be
\mathbb{WO}= \sum_{\{ijkl \}} (T^i -T^j) (T^k -T^l)= {1\over 2}  \mathbb{M}_{ij} T^i T^j\, .
\label{4}\ee
It appears to have 28 terms of the form $T^iT^j$, however, half of them cancels and we are left with 14 terms.  For example for Cartan-Shouten-Coxeter octonion conventions  \cite{Cartan:1926,Coxeter:1946} 
  \bea
    \label{cw}
 \mathbb{WO} && = \sum_{r=0}^{6}   (T^{r+2}-T^{r+4})(T^{r+5}-T^{r+6})\, .    \eea
 We can see these 7x4  terms in right hand side of eq. \rf{Hamming_CW}. But actually, the formula 
 simplifies to   14 terms
\be
\mathbb{WO}=   - \sum_{r=1}^{7}    T^{r}( T^{r+1}- T^{r+2})\, . 
\ee
 Explicitly the 14 terms are 
    \bea
    \label{cwShort}
 \mathbb{WO}=  && - 
 \Big (  T^1(T^2- T^3) + T^2(T^3-T^4) + T^3(T^4-T^5) +T^4(T^5-T^6)
\cr 
\cr
 &&+ \, T^5(T^6-T^7) + T^6(T^7-T^1) +T^7(T^1-T^2)\Big)\, . 
   \eea
   The set of 7 terms in the superpotential in the form \rf{cw}  is easy to compare with 7 octonion associate triads, with 7 quadruples and with  7 codewords  of the cyclic (7,4) Hamming error correcting code.   We show this relation in eq. \rf{Hamming_CW}.

  \be
{\rm \bf Triads }\hskip 1.2 cm {\rm \bf Codewords \hskip 0.7 cm     Quadruples  \hskip 0.5 cm \Rightarrow \hskip 0.3 cm \mathbb{WO}}  \nonumber \ee
\be
\mathbb{WO} = \left(\begin{array}{c|ccccccc|c|c}(137)& \hskip 1 cm1 & 0 & 1 & 0 & 0 & 0 & 1& \hskip 1 cm (2456) & \hskip 1 cm (T^2-T^4)(T^5-T^6) \\(241)&\hskip 1 cm1 & 1 & 0 & 1 & 0 & 0 & 0 & \hskip 1 cm (3567)& \hskip 1 cm(T^3-T^5)(T^6-T^7) \\(352)&\hskip 1 cm 0 & 1 & 1 & 0 & 1 & 0 & 0& \hskip 1 cm(4671)&\hskip 1 cm (T^4-T^6)(T^7-T^1)\\(463)& \hskip 1 cm0 & 0 & 1 & 1 & 0 & 1 & 0 & \hskip 1 cm (5712)& \hskip 1 cm(T^5-T^7)(T^1-T^2)\\(574)& \hskip 1 cm 0 & 0 & 0 & 1 & 1 & 0 & 1& \hskip 1 cm(6123)& \hskip 1 cm(T^6-T^1)(T^2-T^3) \\(615)& \hskip 1 cm1 & 0& 0 & 0 & 1 & 1 & 0& \hskip 1 cm(7234)& \hskip 1 cm(T^7-T^2)(T^3-T^4) \\(726)& \hskip 1 cm0 & 1 & 0 & 0 & 0 & 1 & 1& \hskip 1 cm(1345)&\hskip 1 cm (T^1-T^3)(T^4-T^5) \end{array}\right)
\label{Hamming_CW}\ee

Let us show how the octonion triads are represented in the oriented Fano plane. Each of the 7 lines has 3 points, the arrows show the order, with possible cyclic permutations. For example the first one in eq. \rf{Hamming_CW} is 137, we see it as the internal line going up and to the right, it shows 371. The next one is 241, it is a set of points on a circle. One more, 352 is the one at the bottom, going to the left, it shows as 523, etc.

\begin{figure}[H]
\centering
\includegraphics[scale=0.5]{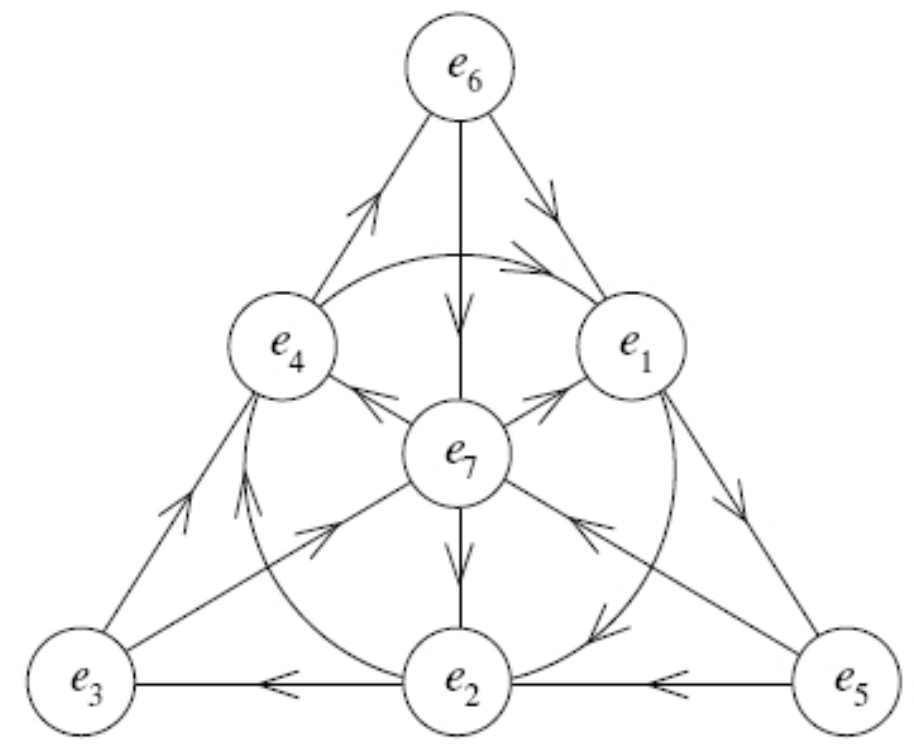}
\caption{\footnotesize  An oriented Fano plane,   Fig. 1 in \cite{Baez:2001dm}. On each of the 7 lines (including the circle) there are 3 points e.g. 1,2, and 4 on a circle. The octonian multiplication rule is build  into the Fano table. For example, one can see from the oriented circle that $e_1\cdot e_2= e_4$.}
\label{fig:Fano1}
\end{figure}

We  studied Minkowski vacua in 7-moduli models with octonionic superpotentials \rf{4}.
We found that   that these  models  have a Minkowski
 minimum at
\be\label{sol}
T^1=T^2=T^3=T^4=T^5=T^6=T^7 \equiv T
\ee
with one flat direction.  

There are 480 different octonion conventions. We have presented a general formula of the superpotential for any octonion convention in \cite{Gunaydin:2020ric}. In all these cases, the matrix $\mathbb{M}_{ij}$ in eq. \rf{4} can be computed either using the general formula or by performing a change of variables. It is therefore not surprising that the eigenvalues of these matrices for all possible octonions are  always the same. We will discuss these eigenvalues and their relation to 3x3 octonionic Hermitian matrices, and to black hole entropy in the next section. 

  We can cut from the superpotential \rf{cw} some terms according to the rules specified via error correcting codes. 
 \begin{figure}[h!]
\centerline{ \epsfxsize 5.5 in\epsfbox{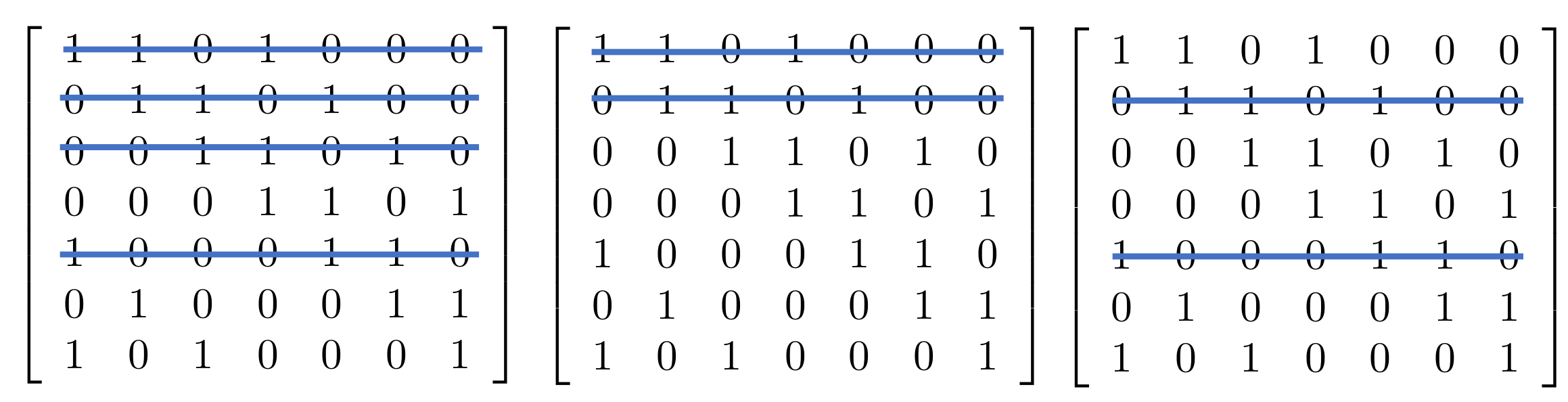}}
\caption{Three sets of codewords with some terms excluded: they codify the superpotentials $\mathbb{WO}_{m,n} $ where the corresponding terms in eq. \rf{cw} are absent. These superpotentials lead to models with Minkowski vacua with two flat directions.}
\label{Split_Codes}
\end{figure} 
The related kinetic terms for the inflaton fields corresponding to all these models with one and two flat directions are
\be
K= - m \log\left(  T_{(1)} + \overline{T}_{(1)}\right) -  n \log\left(  T_{(2)} + \overline{T}_{(2)}\right)
\ee
with $m+n=7$ and cases like $m=0, n=7; m=1, n=6; m=2, n=5; m=3, n=4$. The superpotentials $\mathbb{WO}_{m,n} $  at the vacuum have the following properties:   
\be
  \mathbb{WO}_{m,n} =0\, , \qquad \partial _i  \mathbb{WO}_{m,n} =0
  \ee
  at 
\be
T^1= \dots =T^m\equiv T_{(1)}, \quad T^{m+1}= \dots =T^n \equiv T_{(2)}\, . 
\ee
Based on these M-theory Minkowski vacua we have build $\cN=1$ supergravity  phenomenological models with the potential 
\begin{equation}
V=F (T, \bar T) \Big (1+  \frac{| \mathcal{W}^{\rm oct}|^2}{W_0^2}\Big )+\sum_{i=1}^7 (T^i+\bar{T}^i)^2|\partial_i    \mathcal{W}^{\rm oct}|^2 
\label{generalO}
\end{equation}
where
\be
\mathcal{W}^{\rm oct}\equiv {1\over \sqrt{\prod_{i=1}^7 (2 \, T^i)}}  \mathbb{WO}\, . 
\ee
Along the supersymmetric Minkowski flat directions  we have $\mathcal{W}^{\rm oct} = \partial_i  \mathcal{W}^{\rm oct} = 0$.
 Therefore the full expression for the inflaton potential, for example in the simpletst T-models,  is given by an inflationary  potential for the $\alpha$-attractor models and a cosmological constant
\begin{equation}
V=  \Lambda + m^2 \tanh^2 \sqrt{1\over 6\alpha}\,  \phi \, . 
\label{generalinfl1}
\end{equation}
 Inflation along various flat directions with these kinetic terms leads to  $3\alpha = 7,6,5,4,3,2,1$ and therefore to 7 possible values of the tensor to scalar ratio $r = 12\alpha/N_{e}^{2}$  in the range  $10^{-2}\gtrsim r \gtrsim  10^{-3}$, which should be accessible to  future cosmological observations. They are shown by 7 purple lines  in Fig. \ref{fig:LB} here, taken from  the LiteBIRD satellite mission forecast.

 \section{Properties of the mass matrix in octonion cosmological models}
 
The  octonion superpotentials for models in \cite{Gunaydin:2020ric} with $G_2$ holonomy and 7 moduli have  14 terms   in the  form
\be
 \mathbb{WO}={1\over 2}  \mathbb{M}_{ij} T^i T^j\, . 
\label{octW} \ee
The matrix  $ \mathbb{M}_{ij}$  for the simplest case $\mathbb{WO}$ for Cartan-Shouten-Coxeter  octonion notations is
\be \mathbb{M}_{ \mathbb{WO}} =  \left(\begin{array}{ccccccc} 0&-1&1&0&0&1&-1\\
-1&0&-1&1&0&0&1\\
1&-1&0&-1&1&0&0\\
0&1&-1&0&-1&1&0\\
0&0&1&-1&0&-1&1 \\
1&0&0&1&-1&0&-1\\
-1&1&0&0&1&-1&0
\end{array}\right)
\label{matrixWCSC}\ee
 One can see that it has the property
 $ M_{ii} = \sum_j M_{ij} =0\, ,    \forall i$. In Minkowski  vacuum with $ \mathbb{WO}= \mathbb{WO}_{,i}=0$ the fermion mass matrix is 
\be
{1\over 2} e^{K\over 2}  \chi^i  \mathbb{M}_{ij} \chi^j\, . 
\label{fm}\ee
 The non-vanishing 6 eigenvalues of the $ \mathbb{M}$  matrix, defining the fermion mass eigenstates in Minkowski vacua    solve a double set of cubic equations
\bea
x^3-7x-7=0\, , \qquad  y^3-7y-7=0\, . 
\label{cubic}\eea
The eigenvalues of the fermion mass matrix are
\be \mathbb{M}_{ \mathbb{WO}}^{\rm EV}=   \left(\begin{array}{ccccccc} 
x_1&0&0&0&0&0&0\\
0&x_2&0&0&0&0&0\\
0&0&x_3&0&0&0&0\\
0&0&0&y_1&0&0&0\\
0&0&0&0&y_2&0&0 \\
0&0&0&0&0&y_3&0\\
0&0&0&0&0&0&0
\end{array}\right)
\label{matrixEV}\ee
where  $x_a=y_a$ with $a=1,2,3$ are solutions of the cubic eqs. \rf{cubic}.  
Numerically this gives for a set of $x_1, y_1;\, \, x_2, y_2;\, \,  x_3, y_3$ and a massless one, the following values
\be
3.0489, \, 3.0489; \, \, -1.6920, \, -1.6920; \, \, -1.3569, \, -1.3569; \, \, 0 
\label{fermion}\ee
as shown in \cite{Gunaydin:2020ric}. It looks like the numerical sum of all 3 eigenvalues  vanishes 
\be
3.0489- 1.6920- 1.3569 \approx 0 \, .
\ee
Meanwhile,  can also solve eqs. \rf{cubic} analytically. With $z_a=x_a=y_a$ and $\theta= Arc \tan 3^{-3/2}$ the solutions are
\bea
z_1&=& 2\sqrt{7\over 3}\,  {\rm Re} \,   e^{i {\theta\over 3}} \cr
z_2&=& 2\sqrt{7\over 3}\,  {\rm Re} \,   e^{i {\theta+2\pi\over 3}}\cr
z_3&= &2\sqrt{7\over 3}\,  {\rm Re} \,   e^{i {\theta+4\pi\over 3}}
\eea
Also one finds that
\be
z_1+z_2+z_3=0 \ ,
\ee
which means that indeed the sum of the 3 eigenvalues of the fermions mass matrix vanishes exactly. A number of other relations can be seen in the exact solution:
\bea
z_1z_2+z_2 z_3 +z_1 z_3&=&-7\cr
z_1z_2 z_3&=& 7\cr
z_1^2+ z_2^2+ z_3^2 &=& 2\cdot 7\cr
z_1^3+ z_2^3+ z_3^3 &=& 3\cdot 7\cr
z_1^4+ z_2^4+ z_3^4 &= &2\cdot 7^2\cr
z_1^5+ z_2^5+ z_3^5 &=& 5\cdot 7^2\, . 
\eea
We can compare the eigenvalues of the 3x3 part of the fermion mass matrix with the eigenvalues of the 3x3 octonionic Hermitian matrix studied in \cite{Ferrara:1997uz,Ferrara:2006xx}  which defines the supersymmetric  black hole entropy in 5d. This entropy  was shown in \cite{Ferrara:1996um} to be equal to a square root of the cubic invariant $I_3$ of  $E_{6(6)}$. In \cite{Ferrara:1997uz,Ferrara:2006xx} it was shown how this cubic  invariant  is related to the Jordan algebra $J_3^{\cal O}$ of the 3x3 hermitian matrices over the composition algebra of octonions ${\cal O}$.

A generic element $J$ of $J_3^{\cal O}$ has the form
 \be
J= \left(\begin{array}{ccc}\alpha_1 &o_3 & o_2^* \\o_3^* & \alpha_2 & o_1 \\o_2 & o_1^* & \alpha_3\end{array}\right)
\label{element} \ee
 where $\alpha_a$ are real numbers and $o_a$ with $a=1,2,3$ are elements of ${\cal O}$.  The automorphism of the split  exceptional Jordan algebra is the non-compact $F_{4(4)}$  group. In case of the non-split octonions the automorphism group is $F_4$. An element of $J_3^{\cal O}$ can be brought to a diagonal form by an $F_4$  rotation \cite{Gunaydin:1978jq,Dundarer:1983fe,Ferrara:1997uz,Dray:1999xe}. In case of the black holes  the generic element of $J$ has eigenvalues $\lambda _a$, $a=1,2,3$ and the cubic norm of  $J_3^{\cal O}$ is, as shown in  \cite{Ferrara:1997uz}.
 
 In case of non-split octonions, one also start with the element \rf{element} and diagonalize it using $F_4$ transformation  \cite{Gunaydin:1978jq}. The 3 eigenvalues in this case were shown to satisfy certain  characteristic cubic equation \cite{Dray:1999xe} 
 \be
-\det  (J-\lambda I) = \lambda^3 - (\Tr J)\,  \lambda^2 + \Tr(J\times J) \, \lambda - (\det J) I=0
 \ee
 where in notation of \cite{Dundarer:1983fe}
 \be
J\times J= J^{-1} \det J\, . 
 \ee
In our cosmological model the analogous cubic equation $x^3-7x-7=0$ corresponds to the choice 
\be
\Tr \,  J=0\, , \qquad \Tr \,  J^{-1}=- I\, ,  \qquad \det \, J=7
\ee  
 The choice $\Tr \,  J=0$ according to \cite{Dundarer:1983fe} means that our matrix \rf{element} depends only on 26 parameters and therefore it is a 26-dimensional representation of $F_4$. It is also explained there that the invariants of $F_4$ are
 \be
 \Tr \, J , \qquad  \Tr (  J\times J) \, , \qquad \det J\, .
 \ee
Thus we find that our fermion mass matrix eigenvalues are defined by a cubic equation $x^3-7x-7=0$ of the kind which defines the exceptional Jordan matrix eigenvalues \cite{Dray:1999xe} with special $F_4$ invariant properties.
\be
 \Tr \, J =0, \qquad (J\times J)= -7 \, , \qquad \det J=7\, . 
\label{F4} \ee
  Meanwhile, the relation between the det of the fermion mass matrix   $\mathbb{M}_{ \mathbb{WO}}$ and black hole entropy in the diagonal basis $\sqrt I_3$ is
  \be
\det \mathbb{M}_{ \mathbb{WO}}=  x_1 x_2 x_3    \qquad I_3= \lambda_1\lambda_2 \lambda_3= p^1p^2p^3\, . 
  \ee
In 5d black holes the values of  magnetic charges, $p^1, p^2, p^3$ are less restricted, they do not satisfy a cubic equation of the kind $x^3-7x-7=0$. In fact the entropy of 4d STU black holes we started with in eq. \rf{ww} is the same as the one in 5d under condition that $p^0=q_i=q_2=q_3=0$.

Thus, in addition to numerous relations between various BPS and non-BPS black holes, we have observed here an interesting relation to octonion based cosmological models and fermion mass matrix. 
\section{Discrete symmetry of fermions in  cosmological models }

\begin{figure}[H]
\centering
\includegraphics[scale=0.48]{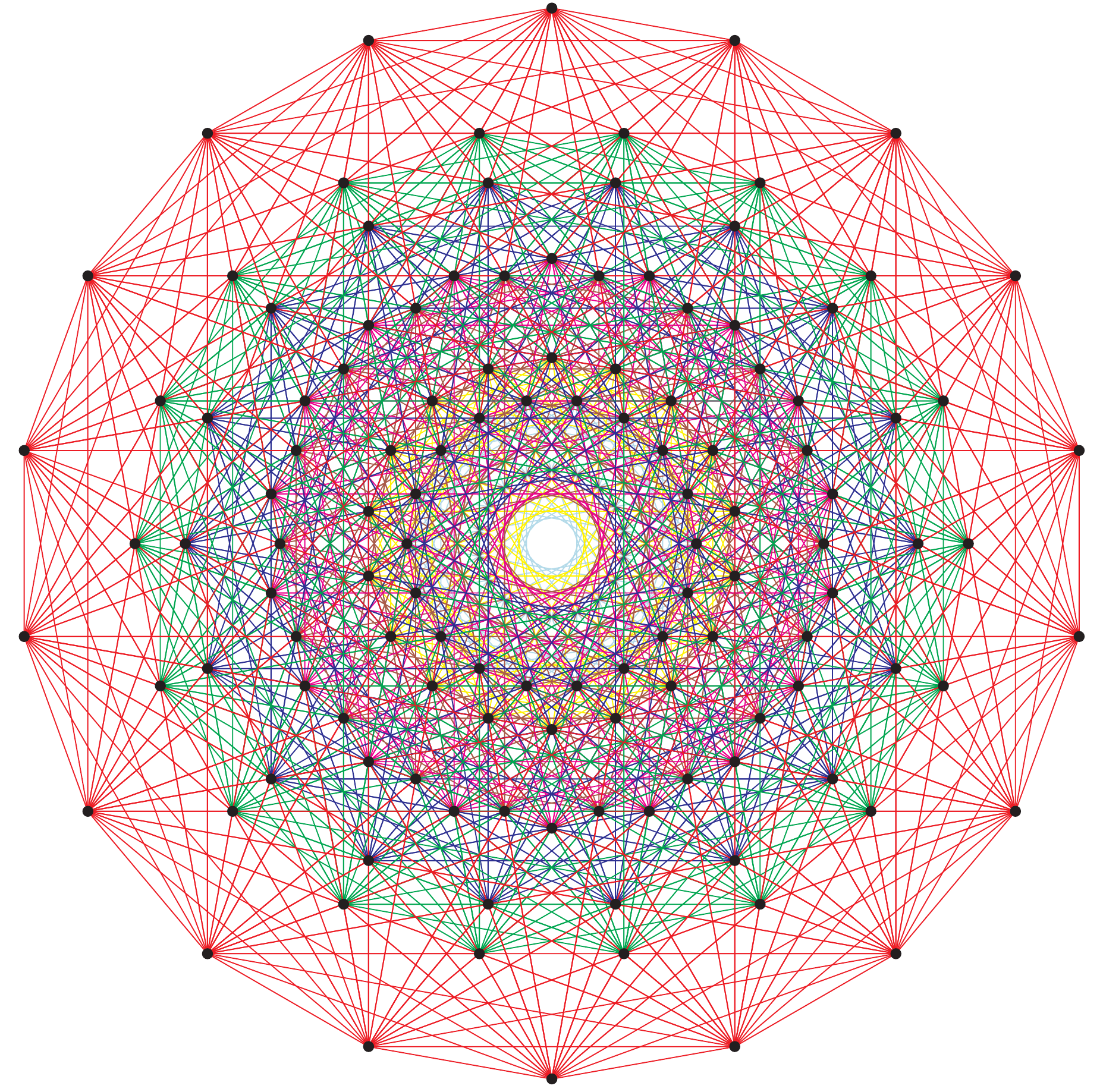}
\caption{\footnotesize  This is the  root system of the Weyl group of $E_7$  projected  into the Coxeter plane as given by 
\href{http://www.math.lsa.umich.edu/~jrs/coxplane.html} {John Stembridge}. The Lie group $E_7$ has a root system of 126 points in 7-dimensional space. One can see these 126 points as 7 groups of 18 points. These 126 points are tightly packed together and this configuration has a total of 2,903,040 symmetries. }
\label{fig:E7roots1}
\end{figure}

\begin{figure}[H]
\centering
\includegraphics[scale=0.465]{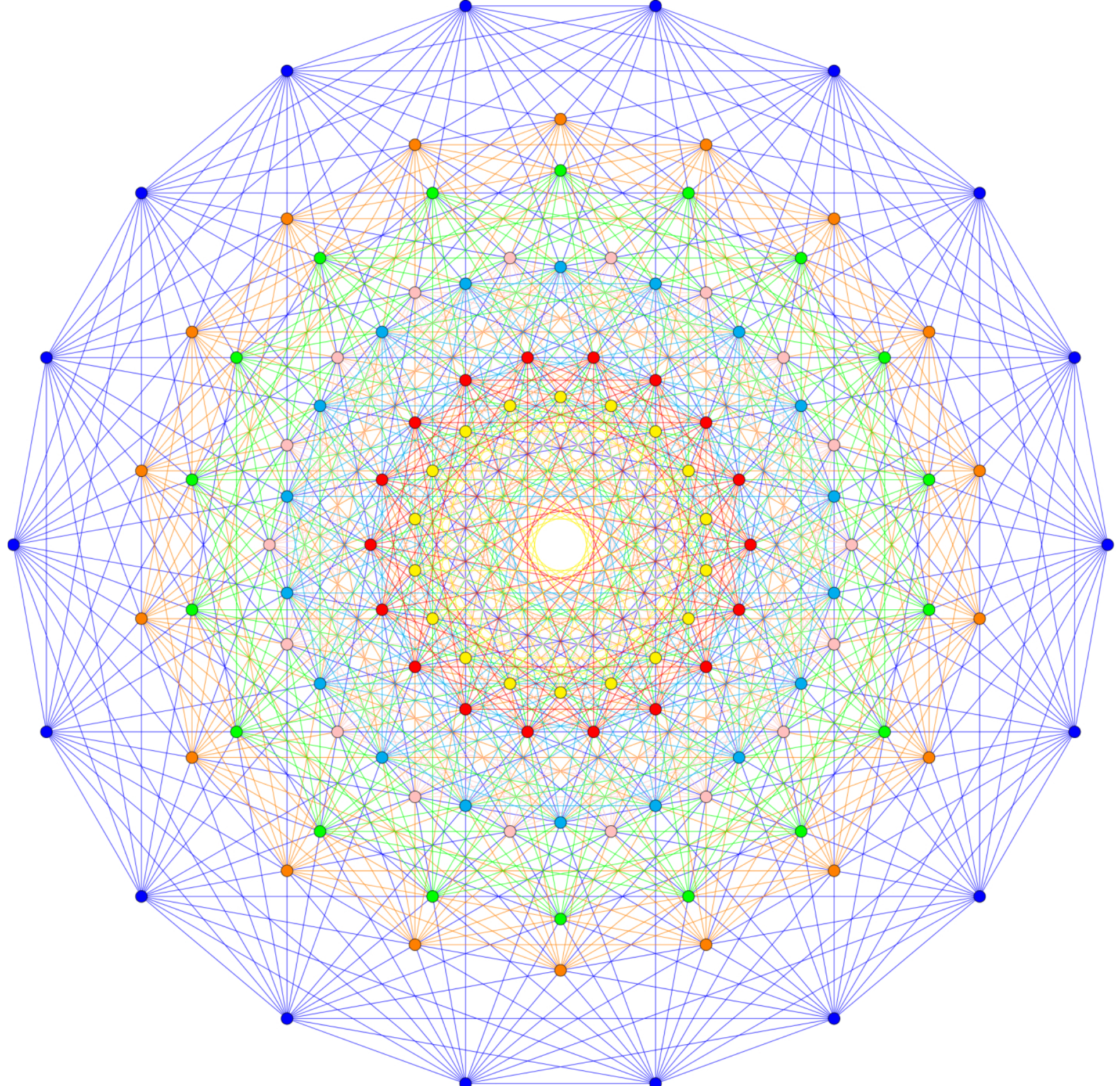}
\caption{\footnotesize  The  Coxeter projections of all exceptional root systems are given by \href{https://tamasgorbe.wordpress.com/2015/10/28/coxeter-projection-of-exceptional-root-systems/} {Tom\'as G\"orbe}, including the $E_7$ case shown here. As in Fig. \ref{fig:E7roots1} we can see 7 circles with  18 points each, a total of 126 points representing a root system of $E_7$.
}
\label{fig:E7roots2}
\end{figure}

The fermion mass matrix  in eq. \rf{fm} 
 at Minkowski vacua in cosmological models \cite{Gunaydin:2020ric} can be brought to a diagonal form  as shown in eqs. \rf{matrixEV}, \rf{fermion}. Since it is a 7x7  matrix, its eigenvalues are invariant under the $O( 7)$  symmetry and its subgroups. The discrete subgroup of it is the Weyl group $W(E_7)$. It is  isomorphic to a finite subgroup of O(7) which is the direct product $Z_2 \times SO_7(2)$. The group $ SO_7(2)$  is the adjoint Chevalley group of order 1,451, 520. The Weyl group $W(E_7)$ has 2,903,040 symmetries.  
 The root system of the Weyl group $W(E_7)$  cannot be visualized since it is an object in 7 dimensions, but the 2-dimensional projections of them,  the Coxeter planes, are well known. We present them in Figs. \ref{fig:E7roots1},  \ref{fig:E7roots2}. 
 
 However, the Weyl group $W(E_7)$ does not preserve the octonion algebra. When one imposes the invariance of the octonion algebra on the transformations of the $E_7$ roots one obtains a finite subgroup of $G_2$, as expected, the adjoint Chevalley group $G_2(2)$ of order 12,096. 
\noindent We now review the analysis of $E_7$ roots and its $G_2(2)$ symmetry following \cite{Koca:1988xe,Karsch:1989gj,Koca:2005fn,Anastasiou:2015ena} and show that it applies to the fermions in cosmological models of \cite{Gunaydin:2020ric}. First we notice that E8 roots can be defined by the integral octonions of the following form.  We take  Cartan-Shoten-Coxeter octonion conventions, which we used in eq. \rf{Hamming_CW}. The triples are 124,235,346,457,561,672,713 and the quadruples are 3567, 4671, 5712, 6123, 7234, 1345, 2456. The set of 240 integer octonions is
 \bea 
\label{16}&&\pm 1, \pm e_i \\ 
\cr \label{triples}
&& {1\over 2} (\pm1\pm e_i\pm e_j\pm e_k)   \\
\cr \label{quad}
 &&{1\over 2} (\pm e_i\pm e_j\pm e_k\pm e_l)\, . 
\eea
Here in \rf{triples} $ijk$ belong to triples, so we have 7x16 = 112 and in \rf{quad} $ijkl$ belong to complementary quadruples, so we have again  7x16 = 112. To this we add 16 from eq.  \rf{16}. This gives the total of 240 integral octonions, which make the E8 roots, also called Cayley integers or octavians. 
From the set of integral octonions above we keep only the ones in 
 \bea 
\label{E7} \pm e_i, \qquad {1\over 2} (\pm e_i\pm e_j\pm e_k\pm e_l)\, . 
\eea
There are 14+7x16=126 integral octonions.
It was shown in  \cite{Koca:2005fn} that
the set of transformations which preserve the octonion algebra of the root system of E7 in \rf{E7} is the adjoint Chevalley group $G_2(2)$. It is possible to decompose these 126 imaginary octonions into 18 sets of 7 imaginary octonionic units that can be transformed to each other by the finite subgroup of matrices. These lead to 18 sets of 7 which we see in Figs. \ref{fig:E7roots1}, \ref{fig:E7roots2}.

Thus it appears that the cosmological models in  \cite{Gunaydin:2020ric} derived from compactification of 11d supergravity on a manifold with $G_2$ holonomy, have some hidden $E_7$ symmetry. It would be nice to understand a relation of all this to the 
 maximal  4d supergravity with \E\, symmetry.

\section{Discussion}
The entanglement of 7 qubits (Alice, Bob, Charlie, Daisy, Emma, Fred and George) was important in M. Duff's studies of black holes in M-theory. In the context of M-theory cosmology, it is not surprising that the concept of 7 qubits and the related tools like octonions, $G_2$ symmetry, Fano Planes, (7,4) Hamming correcting codes also are playing an important role. It would be interesting to develop more understanding of both black holes and cosmological models in M-theory and of the role of octonions in physics. 

Neutrino physics may also require some new ideas to satisfy the current and future data. It was advocated in \cite{Ramond:2020dgm,Perez:2020nqq} that some discrete subgroups of $G_2$, like ${\cal PSL}_2(13)$ might be useful for this purpose.

A nice feature of our cosmological models in \cite{Gunaydin:2020ric} is that they describe a case of maximal supersymmetry spontaneously broken down to a minimal supersymmetry.  These models  will be tested by the future cosmological observations as we show in Figure. \ref{fig:LB}. The most recent forecast of the CMB-S4 in \cite{Abazajian:2020dmr} suggests that the ground based Stage-4 experiments will achieve the science goals of detecting primordial gravitational waves for $r  > 0.003$ at greater than $5\sigma$, or, in the absence of a detection, of reaching an upper limit of $r < 0.001$ at 95\% CL. Therefore  the benchmark targets  of cosmological models in \cite{Gunaydin:2020ric} will be tested during the next decade or two.

\section*{Acknowledgement}
I am grateful to  G. Dall'Agata, M. Duff, S. Ferrara,  S. Kachru, A. Linde,  A. Van Proeyen and Y. Yamada  for stimulating discussions.  
I am  especially grateful to M. Gunaydin for clarification of the  relevant  issues with octonions and exceptional groups.
I am supported by SITP and by the US National Science Foundation Grant  PHY-1720397, and by the  Simons Foundation Origins of the Universe program (Modern Inflationary Cosmology collaboration).
\bibliographystyle{JHEP}
\bibliography{lindekalloshrefs}
\end{document}